\documentclass[a4paper,twocolumn,11pt,accepted=2024-07-30]{quantumarticle}
\pdfoutput=1
\usepackage[numbers,sort&compress]{natbib}
\usepackage[utf8]{inputenc}
\usepackage[english]{babel}
\usepackage[T1]{fontenc}
\usepackage{enumerate}
\usepackage{graphicx}
\usepackage{caption}
\usepackage{subcaption}
\usepackage{dcolumn}
\usepackage{bm}
\usepackage[unicode, colorlinks=true,linkcolor=blue,citecolor=blue,urlcolor=blue]{hyperref}
\usepackage{amsthm}
\usepackage{amsmath}
\usepackage{algorithm2e}
\RestyleAlgo{ruled} 

\usepackage{dsfont}
\usepackage{xcolor}

\usepackage{braket}
\newcommand{\tr}{\mathrm{tr}}
\usepackage{csquotes}
\usepackage{bbold}

\begin{document}

\title{Certification of quantum state functions under partial information}

\author{Leonardo Zambrano}
\email{leonardo.zambrano@icfo.eu}
\affiliation{ICFO - Institut de Ciencies Fotoniques, The Barcelona Institute of Science and Technology, 08860 Castelldefels, Barcelona, Spain}
\orcid{0000-0001-7070-1433}
\thanks{these authors contributed equally to this work.}
\author{Donato Farina}
\email{donato.farina@unina.it}
\affiliation{ICFO - Institut de Ciencies Fotoniques, The Barcelona Institute of Science and Technology, 08860 Castelldefels, Barcelona, Spain}
\affiliation{Physics Department E. Pancini - Università degli Studi di Napoli Federico II, Complesso Universitario Monte S. Angelo - Via Cintia - I-80126 Napoli, Italy}
\orcid{0000-0002-7248-664X}
\thanks{these authors contributed equally to this work.}
%
\author{Egle Pagliaro}
\affiliation{ICFO - Institut de Ciencies Fotoniques, The Barcelona Institute of Science and Technology, 08860 Castelldefels, Barcelona, Spain}
\author{M\'{a}rcio M. Taddei}
\affiliation{ICFO - Institut de Ciencies Fotoniques, The Barcelona Institute of Science and Technology, 08860 Castelldefels, Barcelona, Spain}

\author{Antonio Ac\'{\i}n}
\affiliation{ICFO - Institut de Ciencies Fotoniques, The Barcelona Institute of Science and Technology, 08860 Castelldefels, Barcelona, Spain}
\affiliation{ICREA - Instituci\'{o} Catalana de Recerca i Estudis Avan\c{c}ats, 08010 Barcelona, Spain}

\begin{abstract}
Convex functions of quantum states play a key role in quantum physics, with examples ranging from Bell inequalities to von Neumann entropy. However, in experimental scenarios, direct measurements of these functions are often impractical. We address this issue by introducing two methods for determining rigorous confidence bounds for convex functions based on informationally incomplete measurements. Our approach outperforms existing protocols by providing tighter bounds for a fixed confidence level   and number of measurements. We evaluate the performance of our methods using both numerical and experimental data. Our findings demonstrate the efficacy of our approach, paving the way for improved quantum state certification in real-world applications.
\end{abstract}

\maketitle

{
\section{Introduction}

The value attained by a convex function of a quantum state is generally connected to key concepts in quantum physics~\cite{nielsen2010quantum}.
In experimental contexts, though, these values are often not accessible through direct measurements.
Even for a linear function of the quantum state, i.e.\ the expectation value of a generic observable, measurements can be challenging if the observable is highly nonlocal (e.g., a projector on a highly entangled subspace) or requires measurement devices unavailable in experimental setups.
However, it is possible to achieve an indirect estimation of these observables through the measurements of other more accessible quantities.
Parallel measurements of local Pauli operators constitute a typical example of feasible measurements for a multiqubit system.
A full and perfect tomographic reconstruction, ideally attainable by means of parallel measurements, would naturally lead to the knowledge of any function of the quantum state \cite{james2001measurement, thew2002qudit}. 
Nevertheless, this is unrealistic for two main reasons. 
First, even considering the case of measuring all the possible multiqubit Pauli strings, such measurements contain an inherent uncertainty due to finite statistics effects, and it is not always clear how to translate this uncertainty to the estimation of the convex function. Second, for large systems a tomographically complete set of measurements is exponentially costly both in time and in experimental resources, hence, infeasible in practice. As a result, it is of utmost importance to be able to reliably estimate the value of  convex functions through an incomplete set of measurements, as, for instance, those in compressed sensing quantum tomography \cite{gross2010quantum}. 
That is, one aims at deriving a confidence interval, i.e.\ determining a range of feasible values, for the convex function value given the acquired partial information on the state. \\
The problem, in general, is acquiring increasing interest during the last few years as is related to identifying confidence sets for quantum state tomography \cite{christandl2012reliable, wang-renner-2019, guta2020fast, goh2019experimental,eisert2020quantum, de2023comparison, steinberg2023certifying,norkin2023reliable, de2023user}. These sets are regions within the space of density matrices that contain the experimental state with a certain confidence level. They allow us to establish rigorous bounds for convex functions applied to the state using convex optimization techniques.  However, prior research primarily concentrates on scenarios that require informationally complete measurements, obtainable only through a large number of measurement settings.\\
The size of the confidence region depends on the desired confidence level and the number of available copies of the experimental state. Since quantum resources are limited, it is critical to devise protocols that use the fewest possible copies of the state  for a given confidence level. Here, we present a method to determine the confidence interval of convex functions using informationally incomplete measurements. We demonstrate that, with a fixed confidence level and number of shots, our method yields tighter confidence intervals compared to other previous protocols.\\
To achieve this objective, we present two approaches: 
the first is based on the knowledge of a specific number of expectation values of Hermitian operators (e.g.\ multi-qubit Pauli operators) and assigns confidence intervals to each of these values.
The second is a joint approach that considers general measurements, defined by positive operator-valued measures (POVM). Depending on the type of states that we are considering one method might perform better than the other. 
Both provide useful guidance for certification in realistic experimental settings depending on the input resources.\\
This article is organized as follows.
In Sec.\,\ref{sec:prelim} we introduce the certification scheme in the ideal scenario and the problems arising when dealing with experimental data.
Our two methods are presented in Sec.\,\ref{sec:methods}.
Their performances are first tested for Haar pure states' data in Sec.\,\ref{sec:haar} and then for experimental data produced with the \texttt{ibm\_perth} quantum processor in Sec.\,\ref{sec:experim}.
We draw our conclusions in Sec.\,\ref{sec:conc} together with possible future developments.
\section{Preliminaries}
\label{sec:prelim}
\subsection{Certification in ideal scenario}
We first discuss the ideal scenario, namely, in the absence of any statistical and experimental uncertainties. Then, under partial information about the state $\omega$ given in the form of expectation values of a set of observables $\mathcal{I}=\{O_i\}_i$, certifying that the value of a generic real function of the system state, $\mathcal{F}(\omega)$, is within a certain interval can be treated with the following optimization problem:
\begin{eqnarray}
\begin{array}{rl}
\mathcal{F}_{\rm LB(UB)}=\displaystyle & \min_\rho \; (\max_\rho)  \mathcal{F}(\rho)\\
\textrm{s.t.} & \tr({\rho})=1\\
&{\rho} \geq0 \\
&\tr({O}_i{\rho})=\tr(O_i\omega ) \quad\forall i \in \mathcal{I}\,. \\
\end{array}\label{exact-opt}
\end{eqnarray}
This means that any reconstruction $\rho$ compatible with the equality constraints identified by $\mathcal{I}$ will provide a value $\mathcal{F}(\rho)$ which is greater (lower) or equal than the lower (upper) bound $\mathcal{F}_{\rm LB(UB)}$ determined by Problem~(\ref{exact-opt}).
More importantly, we are guaranteed that the performance associated to the actual state $\omega$, being also a feasible state, is within such a range,
\begin{equation}
\mathcal{F}_{\rm LB}\leq \mathcal{F}(\omega) \leq \mathcal{F}_{\rm UB}
\,.
\label{range-theory}
\end{equation}
If $\mathcal{F}$ is linear,
\begin{equation}
\label{L-hermitian}
    \mathcal{F}_\mathrm{lin}(\omega)=\tr(L \omega)\,,
\end{equation}
with $L=L^\dag$,
Problem \eqref{exact-opt} consists in a semidefinite program (SDP). SDPs are very practical, as they can be efficiently solved with low computational power, and do not feature the problem of local minima~\cite{Watrous_notes, Semidefinite}.   
However, data in \eqref{exact-opt} are perfect, in the sense that do not contain uncertainties, a situation that is unrealistic in experimental scenarios.
}
{
\subsection{Dealing with experimental data: relaxation of the equality constraints}
\label{sec:eqrelax}
The main goal of this work is to generalize the certification scheme \eqref{exact-opt} to a generic situation where the data come from an experiment, taking into account statistical errors.
}

With the aim of relying completely on the information extracted from the experimental data,
we shall relax the equality constraints allowing for fluctuations of the order of the standard errors.
As we shall see, in the experimental scenario the range in \eqref{range-theory} does not apply deterministically, but, instead, holds true with a probability $\mathcal{P}$ for which we derive appropriate lower bounds $1-\delta$,
namely
\begin{eqnarray}
&&\mathcal{F}_{\rm LB}\leq \mathcal{F}(\omega) \leq \mathcal{F}_{\rm UB}
\,,\\
\nonumber
&&{\rm with\, probability \,} \, \mathcal{P}\geq 1-\delta
\,.
\end{eqnarray}
\subsection{Operator convex functions}
Before presenting our methods for certification under measurement uncertainties, a comment is in order. As stated below, the results are valid for target functions $\mathcal F$ in general. However, they will be most interesting to apply when $\mathcal F$ is an (operator) convex function, since the corresponding minimization consists in a convex optimization problem. Convex optimization has the advantage of providing certified lower bounds to the function of interest, as well as that of not having local minima~\cite{boyd_convex_2004}. If $\mathcal F$ is \textit{linear}, then both $\mathcal F$ and $-\mathcal F$ are convex, and both upper and lower bounds can be efficiently and certifiably obtained. Given that linear functions are ubiquitous in quantum mechanics, as they represent the expectation values of operators, we shall apply these methods to several linear functions $\mathcal{F}$ of interest. Additionally, we will show an application to the nonlinear von Neumann entropy (whose negative is convex).

\section{Bounds for the feasible set}
\label{sec:methods}
We now present our two methods in a general formalism.

\subsection{Individual-constraint method}
\textit{
Consider an unknown density operator $\omega$ describing the state of a system of dimension $d$.
Let $O_i$ be a set of Hermitian operators for $i \in \mathcal{I}=\{1,2,\dots, K\}$ with bounded outcomes in $[-1, 1]$, $o_i$ be the corresponding experimental averages obtained with number of measurement shots $N_i$, and $s_i$ be the associated empirical standard deviations.
Then, if $\mathcal{F}$ 
is a function of the density matrix $\rho$, and $\delta \in (0, 1]$, the solution to the optimization problem}    
\begin{eqnarray}
\label{Local}
\begin{array}{rl}
\mathcal{F}_{\rm LB(UB)}= \displaystyle \min_\rho(\max_\rho) & \mathcal{F}(\rho)\\
\textrm{s.t.} & \tr({\rho})=1\\
&{\rho} \geq0 \\
&\vert \tr({O}_i{\rho})-o_i\vert\leq \epsilon_i \quad\forall i \in \mathcal{I}\,, \\
\end{array}
\end{eqnarray}
\textit{is a lower (upper) bound for $\mathcal{F}(\omega)$ with probability at least $1-\delta$. Here
\begin{eqnarray}
\epsilon_i&:=&\min(\epsilon_i^{\rm (H)}, \epsilon_i^{\rm (EB)})\,, \label{min-epsilon-H-EB}\\
\epsilon_i^{\rm (H)}&:=&\frac{\alpha_{K,\delta}}{\sqrt{N_i}}\,, \quad
\epsilon_i^{\rm (EB)}:= {s_i}\frac{\alpha_{K,\delta/2}}{\sqrt{N_i}}+\frac{7}{3} \frac{\alpha_{K,\delta/2}^2}{N_i-1} \,, 
\label{LocHorEB}
\end{eqnarray}
where $\alpha_{K,\delta}:=\sqrt{2 \ln(2 K/\delta)}$.
}\\
The proof of the above statement is based on the application of the Hoeffding or Empirical Bernstein bounds \cite{mnih2008empirical, maurer2009empirical}, followed by the union bound, analogously to the proof in \cite{qot2020}. It is reported in Appendix \ref{app:alfa-K}. {Note also that outcomes are taken in $[-1, 1]$ without loss of generality, as the previous result can easily be adapted to an arbitrary bounded operator by adjusting its maximum and minimum eigenvalue.}

{\it Discussion.---}
The need to consider both Hoeffding and Empirical Bernstein bounds, Eqs.~(\ref{min-epsilon-H-EB}) and (\ref{LocHorEB}), arises because $\epsilon_i^{\rm (EB)}$ allows for fluctuations up to $\alpha_{K, \delta/2}$ times the standard errors $s_i/\sqrt{N_i}$, with a correction term scaling as $\sim 1/N_i$. 
Neglecting this correction (for sufficiently high $N_i$) assumes a normal distribution for the averages, resulting in a feasible set contained within the one defined by $\epsilon_i^{\rm (H)}$ for a rescaled $\delta$. In such cases, $\epsilon_i^{\rm (EB)}$ usually yields tighter results than $\epsilon_i^{\rm (H)}$.
However, this is not always the case, especially when considering the correction term. Consequently, for not too large $N_i$, $\epsilon_i^{\rm (EB)}$ may lead to looser bounds than $\epsilon_i^{\rm (H)}$.
To obtain a tight general bound, we consider the minimum $\epsilon_i=\min(\epsilon_i^{\rm (H)}, \epsilon_i^{\rm (EB)})$ for each constraint $i$, as indicated in Eq.~\eqref{min-epsilon-H-EB}. For \textit{not
huge} numbers $N_i$, we typically expect the Hoeffding bound to be tighter for random states where most or all standard deviations $s_i$ have non-zero values. However, for specifically chosen states, some of the $s_i$ can be nearly zero, making the $EB$ bound tighter. 
Furthermore, we stress that this method does not assume statistical independence between the several random variables $o_i$, opening the way to a broad spectrum of applications.
{We also mention that, for its derivation, method \eqref{Local} should be compared with the feasible regions presented in the Supplemental Material of \cite{qot2020} (that, however, were not applied to optimization problems). By construction, method \eqref{Local} is {\it tighter} for its selective character in using either Hoeffding's bound or Empirical Bernstein's bound instead of solely Hoeffding's bound.
Remarkably, a further improvement of the individual-constraint method is presented in Appendix 
\ref{app:local-improved} showing the possibility of optimizing the step concerning the application of the union bound.
}

{ 
{\it Example.---} 
Let us consider an $n$-qubit system and let $\{ O_i=\otimes_{j=1}^n \sigma_{i_j} \}_i $ be a set of multiqubit Pauli operators for $i \in \mathcal{I}=\{1,2,\dots, K\}$, $K\leq 4^n-1$, $o_i$ be the experimental estimates of $\tr (\omega O_i)$ (correlators for brevity) obtained with number of measurement shots $N_i$ and $s_i=\sqrt{N_i/(N_i-1)}\sqrt{1-o_i^2}$ be the associated empirical standard deviations.
}

{The first consideration that comes out is the scaling of the confidence ranges in the number of qubits $n$.
The functional form of $\alpha_{K, \delta}$
implies a worst case scenario scaling $\epsilon_i\sim \sqrt{n}$.
}
{
Also, with method \eqref{Local} we account for finite statistics effects, and, in particular, for the fact that correlators that have been estimated with more precision deserve a tighter range of freedom (smaller $\epsilon_i$).
In a parallel measurement framework (see Appendix \ref{app:parallel-meas} for details) such correlators are typically the 1-body ones, for which estimations coming from many different measurement settings are typically available. On the contrary, the most imprecise ones are typically the $n$-body correlators for which only one measurement setting can provide information.
In this sense, we remark that correlators extracted from parallel measurements are \textit{not} statistically independent as the estimation of two different correlators can rely on some common shots. 
This happens when two correlators are related through the following rule: from a higher order correlator, e.g.\ the four body correlator $o_{xxz0y}$ for a system of $5$ qubits, we can obtain a lower order correlator by \enquote{substituting} the label of one or more Pauli operators with the identity, obtaining for instance the two-body correlator $o_{0x00y}$. These two correlators{, namely $o_{xxz0y}$ and $o_{0x00y}$,} share some data in their estimation and therefore are \textit{not} statistically independent.

}

\subsection{Joint-constraint method}
\textit{Consider a density operator $\omega$ on a system of dimension $d$. Let $E = 
\{E_i\}_{i=1}^{m}$ define a POVM applied to the system, with $\vec{p} = \left(tr(\omega E_1), tr(\omega E_2), \ldots, tr(\omega E_{m}) \right)$ the exact measurement probabilities and $\vec{q}$ the empirical estimation of $\vec{p}$ obtained by measuring $E$ across $N$ copies of $\omega$. Let $\mathcal{F}$ be a function of density matrices, and $\delta \in (0, 1]$. Then, the solution to the optimization problem
    \begin{equation}\begin{split}
        \mathcal{F}_{\rm LB(UB)} = \min_{\rho } \; (\max_{\rho}) \quad & \mathcal{F}(\rho)\\
        \mathrm{s.t.} \quad &  \tr(\rho) = 1\\
        &{\rho} \geq 0 \\
        &\sum_{i=1}^m | \mathrm{tr} (E_i \rho) - q_i | \leq \epsilon\ , 
    \end{split}\label{joint_method}\end{equation}
is a lower (upper) bound for $\mathcal{F}(\omega)$ with probability at least $1-\delta$. Here 
\begin{align}\label{eq:bound_joint}
    \epsilon = \sqrt{ \frac{2}{N} \ln \frac{2^m}{\delta}}.
\end{align}}\\
This result is based on the Bretagnolle–Huber–Carol inequality~\cite{bretagnolle1978lois}, that bounds the total variation distance $\| \vec{p} - \vec{q}\|_1$ between a vector of multinomially distributed random variables and its expected values (see Appendix \ref{ap:l1_bound}). The total variation distance has recently been used to define distance measures between quantum states, measurements, and channels based on their statistical distinguishability \cite{ Maciejewski2023operationalquantum}.
\subsection{Comparing the two methods}
With both methods \eqref{Local} and \eqref{joint_method}, we will be interested in the case of the function $\mathcal F$ being linear or convex, as they both lead to convex optimization problems. \\
Let us mention a few differences between methods \eqref{Local} and \eqref{joint_method}. In a typical quantum-state measurement scenario, $\vec q$ contains information on the counts of each possible outcome of every measurement, whereas in \eqref{Local}, the counts are grouped into expectation values of Hermitian operators. Then, the dimension of $\vec q$ is typically larger than the number of Hermitian operators since these operators can be chosen to avoid redundancy. For example, measuring the 9 Pauli basis that are informationally complete for $2$ qubits gives a $\vec{q}$ of dimension $36$ 
(we call
for brevity
\textit{Pauli basis} a particular choice of parallel measurement setting, see Eq.\,\eqref{meas-setts} and related discussions in Appendix \ref{app:parallel-meas}). 
The same information can be grouped into $15$ Pauli operators. Another difference between the two methods is that \eqref{joint_method} bounds the estimation errors jointly in a single equation, with $\epsilon = \sqrt{\frac{2}{N} \ln \frac{2^m}{\delta}}$, while the local method employs one constraint for each Hermitian operator. 
As we shall see later the two methods, beyond in general responding to different experimental requests, have different performances in terms of tightness of the resulting bounds.
However, as a common feature, regarding the behavior in terms of the confidence level, in both methods the errors in the constraints in Eqs. \eqref{LocHorEB} and \eqref{eq:bound_joint}  are monotonically decreasing with increasing $\delta$, as they are proportional to $\sqrt{\ln \delta^{-1} }$.
This implies that having a higher 
confidence level
$1-\delta$ 
results in a decreasing (increasing) behavior for $\mathcal{F}_{\rm LB(UB)}$, i.e.\ 
our bounds become looser but more secure.\\
Recently de Gois and Kleinmann proposed a method to find confidence regions based on the vector Bernstein inequality \cite{de2023user}. Those regions can also be integrated into convex optimization routines to find bounds for convex functions, but require the measurement of an informationally complete POVM. 
We show in what follows that for paradigmatic figures of merit in quantum information our bounds outperform the bounds provided by Ref.\,\cite{de2023user}. 

\section{Numerical simulations}
\label{sec:haar}
\begin{figure}[t]
	\centering
	\includegraphics[width=1\linewidth]{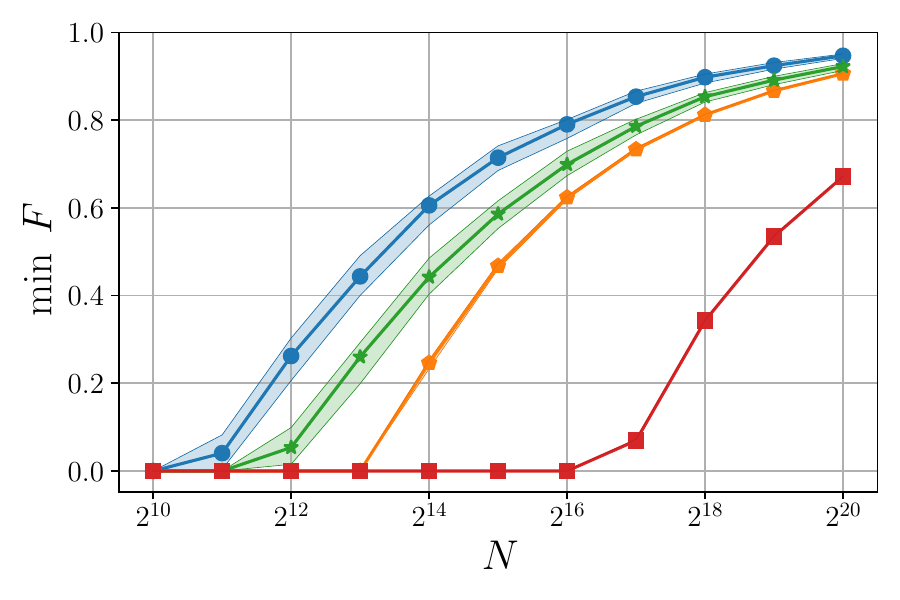}
	\caption{Minimum preparation fidelity of $4$-qubit states as a function of the total number of shots $N$.
 The confidence level is $0.997$. The solid lines represent the median value obtained from $100$ Haar-random pure states using the joint-constraint method (blue dots), individual-constraint method (green stars), method from Ref. \cite{de2023user} (orange pentagons) and {classical shadows (red squares)}, while the shaded areas depict the interquartile ranges.
When applying our two methods we considered $16$ random measurement settings {(more precisely, we draw 16 random settings for each drawn random state)}, while for the method from Ref. \cite{de2023user} all the $81$ measurement settings.
{We remark that $N$ is the total number of shots in the whole simulated experiment, thus, in
general, this implies less shots per measurement setting.}
}
    \label{fig:min_fid2}
\end{figure} 

\begin{figure}[t]
	\includegraphics[width=1\linewidth]{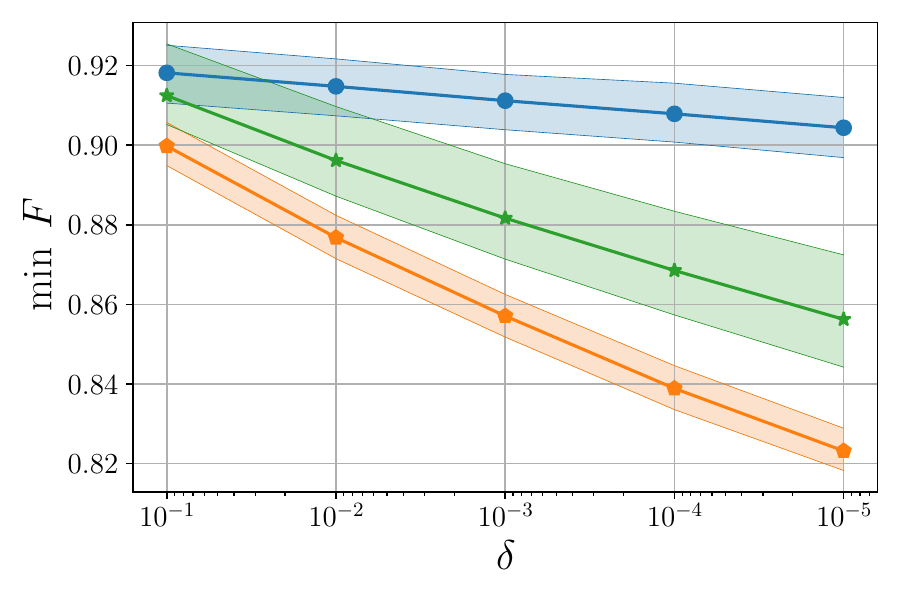}
	\caption{Behaviors in terms of the confidence level.
 We consider all $9$ possible measurement settings for two qubits, a number of shots $N=16379$ ($N\approx 2^{14}$, $\approx 1820$ shots per measurement setting)  and plot the median of the minimum preparation fidelity over $100$ Haar-random pure states as function of decreasing $\delta$. The shadowed region indicates interquartile ranges.
 Different curves and associated shadowed regions refer to different methods. Joint-constraint method (blue dots), individual-constraint method (green stars) and method from Ref. \cite{de2023user} (orange pentagons).
 }
    \label{fig:min_fid_delta}
\end{figure} 
\subsection{Certification of state preparation}
The first application we discuss involves a certification protocol for preparing quantum states. Let us assume our goal is to prepare a pure quantum state $|\psi \rangle$ within a physical system, and we want to check the quality of the preparation, in terms of the worst-case state preparation. To achieve this, we perform measurements on the system and, given this information, minimize the fidelity $F (\rho) = \langle \psi | \rho | \psi \rangle$ between the states $\rho$ in the feasibility set $\Omega$, and the desired state $| \psi \rangle$. Therefore, for an experimental state $\omega$, we obtain the lower bound $\langle \psi | \omega |\psi \rangle \geq \min_{\rho \in \Omega} F (\rho)$.\\
We conducted numerical simulations for this protocol. Specifically, we prepared $100$ Haar-random $4$-qubit pure states $|\psi \rangle$ and we measured them using  $16$ randomly chosen Pauli bases,  evenly distributing the $N$ copies of the state. 
Subsequently, we apply our two protocols and protocol A from Ref. \cite{de2023user} for a selected confidence level. In the case of protocol $A$, we measure the $81$ Pauli bases that are informationally complete for $4$ qubits. Finally, we calculated the fidelity of the resulting states with the initially prepared states $|\psi\rangle$. 
{We also applied fidelity estimation using classical shadows with random Pauli bases \cite{huang2020predicting}. The minimum fidelity in this case is given by $\hat{f} - \epsilon'$, where $\hat{f}$ is the classical shadows estimator of the fidelity and $\epsilon' = \frac{64 \times 4^4}{N} \ln \left( 2/\delta \right)$ quantifies the error in the estimation.}

In Figure \ref{fig:min_fid2}, fixing $1-\delta=0.997$,
we show the behavior of the minimum preparation fidelity in terms of the number of shots.
Blue dots and green stars represent the results obtained using the joint-constraint method and individual-constraint method, respectively, while the orange pentagons correspond to the results obtained with the method from Ref. \cite{de2023user} {and the red squares are the results using classical shadows}. 
We observe that, using the same data, both our bounds outperform the bounds provided by Ref. \cite{de2023user} {and classical shadows}.
In particular, 
 the fidelity bounds are significantly tighter when employing the joint-constraint method, which requires approximately four times fewer quantum resources than method from Ref. \cite{de2023user}.
Also the experimental demand is lower, since we only measure around $1/5$ of the total number of settings in our protocols. \\
In Fig.\,\ref{fig:min_fid_delta}, fixing a total of $2^{14}$ shots, we check the scaling of the protocols with the confidence level $1-\delta$ (decreasing values of $\delta$).
Also in this case our two methods provide tighter results with respect to the method from Ref. \cite{de2023user}, with the results obtained from the joint-constraint method being the  tightest and, remarkably, pretty stable with increasing confidence level.

\subsection{Certification of the maximum von Neumann entropy}

\begin{figure}[t]
	\centering
	\includegraphics[width=1\linewidth]{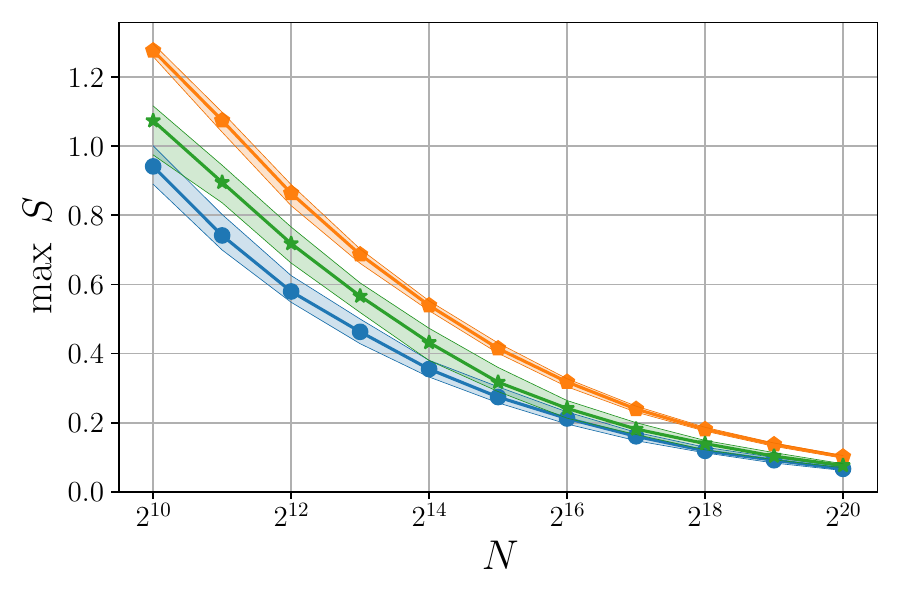}
	\caption{Maximum entropy of $2$-qubit pure states as a function of the total number of shots $N$. The confidence level is $0.997$. The solid lines represent the median value obtained from $100$ Haar-random pure states using the joint-constraint method (blue dots), individual-constraint method (green stars) and method from Ref. \cite{de2023user} (orange pentagons), while the shaded areas depict the interquartile ranges.
 We considered all the $9$ possible measurement settings.}
 \label{fig:max_ent}
\end{figure}
We now aim to determine an upper bound for the von Neumann entropy $S$ of a state $\omega$. This is a convex optimization problem, since $S$ is a concave function. Then, we perform measurements on the system and maximize the function $S(\rho) = - \tr(\rho \ln \rho)$ over all the states $\rho$ compatible with the feasibility set $\Omega$. \\
In Fig \ref{fig:max_ent}, we prepare $100$ Haar-random pure states of $2$ qubits, and we measure them using all $9$ Pauli observables $\{ \sigma^{(1)}\otimes \sigma^{(2)} \}$, with $\sigma = X, Y, Z$. 
The protocols are then applied with confidence level $0.997$. The blue dots and green stars showcase the maximum entropy yielded by methods \eqref{joint_method} and \eqref{Local}, respectively, for different numbers of total shots. 
The orange pentagons are the results using the method A from Ref. \cite{de2023user} with the same data and confidence level.  
The conclusions we can draw are analogous to the ones extracted from the previous analysis.
Both our methods provide tighter results than the method from Ref.\,\cite{de2023user}, with the best performance being reached with the joint-constraint method.

\section{Experimental results}
\label{sec:experim}

\begin{table*}[t!]
\centering
\begin{tabular}{c|ccc|ccc}
\multicolumn{1}{c}{} & \multicolumn{3}{c}{\textbf{Individual-constraint}} & \multicolumn{3}{c}{\textbf{Joint-constraint}} \\
\hline
state & $\min F$ & $\max F$ & $\max S$ & $\min F$ & $\max F$ & $\max S$ \\
\hline
Product &0.916  &0.983  &0.397 & 0.891 & 0.993 & 0.492\\
Bell pair &0.747  &0.862  &0.986 & 0.681 & 0.879 & 1.102 \\
GHZ &0.730  &0.878  &1.143 &  0.671 & 0.910 &  1.289\\
\hline
\end{tabular}
\caption{Analyzing experimental data produced with \texttt{ibm\_perth} quantum processor for $4$ qubits with both individual-constraint method and joint-constraint method for confidence level $0.997$.
Given a selected target state we measured with $16$ different random settings, performing $N=2^{18}$ total number of shots.
The target states are:
Product state,
GHZ, 
Bell pair.
We remark that being in the informationally incomplete scenario method from Ref.\,\cite{de2023user} cannot be applied.
}
\label{table:both}
\end{table*}
We now present results based on actual experimental data. A fundamental difference with respect to the previous analysis is that the target state (the pure state one aims to prepare) and the actual state (the state actually produced in the experiment) do not coincide. We performed experiments using the \texttt{ibm\_perth} quantum processor. For $4$ qubits, we aimed to prepare a random product state, a tensor product of two Bell states, and a GHZ state. We then measured each of these states with a certain number of random Pauli bases, using $2^{14}$ shots per basis. We compared the experimental state with the theoretical ones in terms of maximum and minimum fidelity, and we also calculated the maximum von Neumann entropy. The results using methods \eqref{Local} and \eqref{joint_method} are displayed in Table \ref{table:both}.
We notice that an analysis with the method from Ref.\,\cite{de2023user} is not possible because the data at our disposal are informationally incomplete.

In contrast to the numerical simulations of Sec. \ref{sec:haar}, the best performance is obtained here with the individual-constraint method, i.e., it provides tighter values for maximum and minimum preparation fidelity and maximum von Neumann entropy.\\
{On this regard, we notice that the associated target pure states (i.e. what we wanted to theoretically prepare), may possess some structure, an important difference with respect to random Haar states.
This implies that they could be described in terms of few relevant Pauli strings.
This generally makes the individual-constraint method more efficient  for these \textit{particular}
states, because, at variance with the joint-constraint method, it bounds each Pauli string expectation value independently.}
Furthermore, additional factors should be taken into account when examining the data presented in the table.
Product states are usually prepared with low error in superconducting platforms.
Then, the results for the minimum fidelity are comparable to the value $0.896$ obtained by numerical simulations of the joint-constraint method assuming perfect preparation and a total $N$ of $2^{18}$ shots. 
Worse performances in preparation are instead obtained in the case of entangled target states.
For this reason, when compared with the product state, the results we obtained for minimum and maximum fidelity  are smaller in the cases of tensor product of two Bell states and a GHZ state, and, analogously, for maximum von Neumann entropy the last two state preparations yield larger upper bounds.
\section{Discussion and conclusions}
\label{sec:conc}
We faced the problem of identifying a certified interval for a generic function of a quantum state in the informationally incomplete scenario and under finite statistics effects, focusing on the treatable cases of linear functions and the more general operator convex functions.
On this regard, we introduced two methods that,
considering paradigmatic figures of merit, lead to tighter bounds than the ones extractable from the recent method presented in Ref.\,\cite{de2023user}.\\
We remark that our schemes are general in the sense that they enable us to find rigorous confidence bounds for the experimental values of convex functions of density matrices using various types of measurements.   
This flexibility allows the method to be effectively applied to a wide range of different experimental setups, as researchers can conveniently choose the most suitable measurements for their certification task. \\
Analytical methods like classical shadows \cite{huang2020predicting} or direct fidelity estimation \cite{Flammia2011} often require new calculations to obtain confidence bounds when the function to certify or the measurements to perform are modified. In contrast, using confidence regions, convex optimization can be directly applied without the need for any additional calculations. Then, the method can be straightforwardly applied to numerous other interesting functions, including the quantum Fisher information \cite{MullerRigat2023certifyingquantum}, Bell inequalities, entanglement witnesses, and more. 
We mention that one can use our schemes to find a compressed sensing reconstruction by considering the argument of the minimization for any convex function \cite{kalev2015quantum},
but with rigorous and tight bounds for the constraints provided by our methods.
\\
Furthermore, since finding the quantum state that describes a physical system is not our primary objective, the informational completeness of the measurements is not as crucial as in quantum tomography. In principle, data from any measurement will provide bounds for the convex functions. This contrasts with other results on confidence regions for quantum states, which primarily focus on informationally complete measurements. However, if we intend for the bounds of our method to be non-trivial, we require a reasonable number of settings, as in compressed sensing tomography.  \\
About the joint-constraint method, improvements in confidence regions require tighter bounds for the discrepancy between the experimental and exact probabilities, $|| \vec{p} - \vec{q}||_1$. 
This could be done by trying different concentration inequalities. Furthermore, it is worth considering that the $1$-norm constraint may not always be the best choice. Then, confidence bounds using other $p$-norms could potentially yield better results.
Other concentration inequalities beyond Hoeffding and Empirical Bernstein bounds can also provide improvements in the individual-constraint method by 
adding their resulting errors in the minimization appearing in \eqref{min-epsilon-H-EB}. 
\\
Finally, we notice that the methods can include the case of detection inefficiencies by modifying the POVM elements to include an additional output corresponding to non-detected events, 
 and they might be extended to other estimation tasks, such as, process tomography by means of the Choi-Jamiolkowski isomorphism, or detector tomography. 

\begin{acknowledgments}
This work was supported by the Government of Spain (Severo Ochoa CEX2019-000910-S, FUNQIP, Misiones CUCO Grant MIG-20211005, European Union NextGenerationEU PRTR-C17.I1 and Quantum in Spain), Fundació Cellex, Fundació Mir-Puig, Generalitat de Catalunya (CERCA program), the ERC AdG CERQUTE, the AXA Chair in Quantum Information Science, and the EU PASQUANS2 project Quantera project Veriqtas.
D.F. acknowledges support from PNRR MUR Project No. PE0000023-NQSTI. E.P. acknowledges support from “La Caixa” Foundation fellowship (ID 100010434, code LCF/BQ/DI23/11990078).

\end{acknowledgments}

\bibliographystyle{apsrev4-2}
\bibliography{bib}

\onecolumngrid
\appendix

\section{Proof of the individual-constraint method \eqref{Local}}
\label{app:alfa-K}

The proof follows the lines of that given in Appendix A of \cite{qot2020}, being, however,  valid not only for the Hoeffding bound as in \cite{qot2020}, but also for the Empirical Bernstein bound.

The trace and semipositivity conditions in \eqref{Local} are always obeyed by the actual quantum state $\omega$. Consider $N_i$ shots for the measurement of $\mathbb E(o_i) = \tr(O_i\omega)$ for given $i\in \{1,2,\dots, K\}$. This entails that there are $N_i$ real-valued i.i.d.\ random variables for each $i$. They have, for simplicity, range $R=[-1,1]$, and their empirical average and standard deviation are $o_i$ and $s_i$, respectively.
Given these data, the probability for a single fluctuation to lie outside a range defined by $\epsilon_i$ can in general be upper bounded, 
\begin{equation}
\mathrm{Prob}(| \mathbb{E}(o_i)-o_i | > \epsilon_i)\leq \delta\,,
    \label{chernoff-ineq-1}
\end{equation}
where concentration inequalities \cite{boucheron2013concentration} allow finding
closed expressions for $\epsilon_i(\delta)$.
Defining $\alpha_{\delta}:=\sqrt{2 \ln(2/\delta)}$, then
\begin{eqnarray}
\epsilon_i^{\rm (H)}(\delta)&:=&\frac{\alpha_{\delta}}{\sqrt{N_i}}\ , \label{LocH1}\\
\epsilon_i^{\rm (EB)}(\delta)&:=& {s_i}\frac{\alpha_{\delta/2}}{\sqrt{N_i}}+\frac{7}{3} 
\frac{\alpha_{\delta/2}^2}{N_i-1} \label{LocEB1}
\end{eqnarray}
are eligible options, corresponding to Hoeffding's and Empirical Bernstein's \cite{mnih2008empirical, maurer2009empirical} inequalities, respectively.
This implies that to restrict the set as much as possible we can take
\begin{eqnarray}
\epsilon_i(\delta)=\min[\epsilon_i^{\rm (H)}(\delta), \epsilon_i^{\rm (EB)}(\delta)]\,.
\label{min-epsilon-H-EB1}
\end{eqnarray}
From the union bound, and using $\mathbb E(o_i) = \tr(O_i\omega)$,
\begin{equation}
  \mathrm{Prob}\left[\exists\, i  \ \ s.t. \,| \tr(O_i\omega)-o_i| > 
  \epsilon_i (\delta')
  \right]
  \leq \sum_{i=1}^K
  \mathrm{Prob}[\ | \tr(O_i\omega)-o_i | >\epsilon_i(\delta')] \,,
\end{equation}
we obtain a bound for the violation of any of the $K$ constraints,
\begin{equation}
  \mathrm{Prob}\left[\exists\, i  \ \ s.t. \,| \tr(O_i\omega)-o_i |> 
  \epsilon_i (\delta')
  \right]
  \leq K \delta' \,.
\end{equation}
By rescaling $\delta'=\delta/K$, we conclude that with probability at least $1-\delta$ all $\epsilon$ constraints in \eqref{Local} are fulfilled by $\omega$. As such, with probability at least $1-\delta$ the state $\omega$ is in the feasible space of Problem \eqref{Local}. Since the minimum (maximum) $\mathcal F_{\rm LB(UB)}$ serves as lower (upper) bound to any state in the feasible space, it bounds $\mathcal F(\omega)$ with probability at least $1-\delta$. \qed 

We remark that, as the union bound does not assume statistical independence between the several random variables $o_i$, this implies that the method \eqref{Local} can be applied for this case too.

\section{Improving the individual-constraint method \eqref{Local}}
\label{app:local-improved}
We present here a more involved improved version of the Individual-constraint method \eqref{Local}.
We start noticing that for generic
\begin{equation}
\vec{\delta}=(\delta_1, \delta_2, \dots, \delta_K)^T\,,
\quad
\sum_{i=1}^K \delta_i=\delta\,,
\quad
\delta_i\geq 0\,,
\end{equation}
{  
\begin{eqnarray}
\label{Local-improved}
\begin{array}{rl}
\mathcal{F}_{\rm LB(UB)}(\vec{\delta})= \displaystyle \min_{{\rho}}(\max_\rho)  &\mathcal{F}(\rho)\\
\textrm{s.t.} & \tr({\rho})=1\\
&{\rho} \geq0    \\
&\vert \tr({O}_i{\rho})-o_i\vert\leq 
\min(\frac{\alpha_{\delta_i}}{\sqrt{N_i}}\,,\, {s_i}\frac{\alpha_{\delta_i/2}}{\sqrt{N_i}}+\frac{7}{3} 
\frac{\alpha_{\delta_i/2}^2}{N_i-1} ) \quad\forall i \in \mathcal{I}\,,
\end{array}\hspace{3cm} &&
\end{eqnarray}
is a lower (upper) bound for $\mathcal{F}(\omega)$ with probability at least $1-\delta$.
}

Method \eqref{Local} sets $\delta_i=\delta/K$, but there are in general non-uniform choices of the vector probabilities $\vec{\delta}$ for which method \eqref{Local-improved} is tighter.

\section{Parallel measurements and correlators}
\label{app:parallel-meas}
In introducing the parallel measurement framework we follow, among others, the introduction of Reference \cite{qot2020}.

A common decomposition for the density operator ${\rho}$ of an $n$-qubit system is in terms of Pauli operators, 
\begin{eqnarray}
    {\rho}=
\frac{1}{2^n} \sum_{i=0}^{4^n-1} \tr({O}_i \rho)\, {O}_i\,,
\quad
{O}_i:={\sigma}_{i_1}\otimes
{\sigma}_{i_2}\otimes
\dots\otimes
{\sigma}_{i_n}\,,
\label{multiqubitpauli}
\end{eqnarray}
with 
${\sigma}_{i_j} \in \{{\mathbb{1}}_2, {\sigma}_{x}, {\sigma}_{y}, {\sigma}_{z}\}
$ and 
where the term $\tr({O}_0 \rho)=1$, as ${O}_0:= \otimes_{j=1}^n {\mathbb{1}}_2={\mathbb{1}}_{2^n}$. 
This means that, in principle, in order to unequivocally identify the quantum state, one should measure the $4^n-1$ expectation values of the operators \eqref{multiqubitpauli}.
However, nowadays parallel measurements are experimentally under control.
One can concurrently measure in one shot all the $n$ qubits in different local bases, obtaining $n$ binary outcomes $\vec{a}:=(a_1, a_2, \dots, a_n)$, where $a_j \in \{-1,1\}$, from a given measurement setting
\begin{equation}
\vec{\alpha}:=({\alpha_1},
{\alpha_2}, 
\dots
{\alpha_n})\,,
\label{meas-setts}
\end{equation}
with
${\alpha_j} \in \{{x}, {y}, {z}\}$. Collecting the corresponding counts $N_{a_1, a_2, \dots, a_n\vert \alpha_1, \alpha_2, \dots, \alpha_n}$ leads to the estimated probabilities 
\begin{equation}
p(a_1, a_2, \dots, a_n\vert \alpha_1, \alpha_2, \dots, \alpha_n)\approx\frac{N_{a_1, a_2, \dots, a_n\vert \alpha_1, \alpha_2, \dots, \alpha_n}}{N_{\alpha_1, \alpha_2, \dots, \alpha_n}}
\end{equation}
of obtaining the array of outcomes $(a_1, a_2, \dots, a_n)$ from the measurement setting \eqref{meas-setts}. 
There are in total $2^n$ possible outcomes $(a_1, a_2, \dots, a_n)$ 
and
$3^n$ possible measurement settings \eqref{meas-setts}.

From the above counts $N_{a_1, a_2, \dots, a_n\vert \alpha_1, \alpha_2, \dots, \alpha_n}$ one can infer the correlators $\tr({O_i} {\rho})$, by marginalizing the appropriate qubits.
For instance, in the case of 4 qubits and from the measurement setting $(x,x,y,z)$, one can estimate the correlator
\begin{eqnarray}
tr({\sigma}_{x} \otimes  {\mathbb{1}}\otimes {\sigma}_{y} \otimes {\sigma}_{z}\, {\rho})\approx 
\sum_{a_1, a_2, a_3, a_4} a_1 a_3 a_4
\frac{N_{a_1, a_2, a_3, a_4\vert x, x, y, z}}{N_{x, x, y, z}}
\,.
\label{eq:fromxxyz}
\end{eqnarray}
More generally, a given $k$-body correlator ($k$ counts how many Pauli matrices are not the identity in the string \eqref{multiqubitpauli}) can be estimated from several measurement settings,
\begin{eqnarray}
   \tr(\otimes_{j=1}^n {\sigma}_{i_j}   \, {\rho})\approx o_i =
   \frac{
   \sum_{\vec{\alpha}\vert i}
   \sum_{\vec{a}}
   (\prod_{j\vert i} a_j)
   N_{\vec{a}\vert \vec{\alpha}}}{
   \sum_{\vec{\alpha}\vert i}
   \sum_{\vec{a}}
   N_{\vec{a}\vert \vec{\alpha}}}\,.
   \label{corr-general}
\end{eqnarray}
The $\sum_{\vec{\alpha}\vert i}$ means that the sum must be taken only over the measurement settings $\vec{\alpha}$ that are \enquote{compatible} with the correlator $\tr(\otimes_{j=1}^n {\sigma}_{i_j}   \, {\rho})$. 
More precisely, the number of settings that must be considered for the estimation of a $k$-body correlator is
\begin{eqnarray}
N_{\rm ms}(k)=3^{n-k}\,.
\label{NsettBody}
\end{eqnarray}
Those settings are the ones that can coincide with the correlator string $(i_1,\dots,  i_n)$ when substituting one or more elements of the measurement setting string  $(\alpha_1, \dots, \alpha_n)$ with the identity (the latter being identified by $i_j=0$). The $\prod_{j\vert i} a_j$ imposes instead to consider in the product only the locations $j$ for which $i_j\neq 0$. 
Hence, the denominator in~\eqref{corr-general},
\begin{eqnarray}
   N_i= \sum_{\vec{\alpha}\vert i}
   \sum_{\vec{a}}
   N_{\vec{a}\vert \vec{\alpha}}\,,
   \label{shots-def}
\end{eqnarray}
counts the number of shots used for the estimation of the correlator $i$. According to our notation, $N_0$ coincides with the total number of measurement shots in the experiment and $o_0=1$ accounts for the normalization condition of the density matrix. Furthermore, the above formalism includes the case where not all the settings are measured (informationally incomplete scenario), as in that case one can set to zero the corresponding counts and apply \eqref{corr-general} only when its denominator $N_i$ is non-zero.

\section{Proof of the joint-constraint method \eqref{joint_method}}
\label{ap:l1_bound} 

The proof follows along the lines of Appendix \ref{app:alfa-K}, but with a different concentration inequality. Consider we have $N$ copies of a quantum state $\omega$ and we measure them using a POVM $E$ with $m$ possible outcomes. With $\vec{p} = \left(tr(\omega E_1),tr(\omega E_2), ..., tr(\omega E_{m})  \right)$, let $\vec Z = (Z_1, Z_2, \ldots , Z_m)$ be the outcomes (counts) of that measurement. Now, $\vec Z$ is a multinomially distributed random vector with parameters $(p_1, p_2, \ldots , p_m)$ such that $\sum_{i=1}^m Z_i = N$. The Bretagnolle–Huber–Carol bound~\cite{bretagnolle1978lois} holds and can be written as:
\begin{align}
\mathrm{Prob} \left( \sum_{i=1}^m \left|\frac{Z^{(i)}}{N} -p_i\right|  \geq  \epsilon \right) \leq 2^m e^{-N\epsilon^2/2} =: \delta \ ,\label{eq:BretagnolleBound}
\end{align}
bounding the probability for the total $\ell_1$-norm error to be above $\epsilon$. Identifying $\vec q = \vec Z/N$ and solving for $\epsilon$ as a function of $\delta$, we can state that, with probability at least $1-\delta$, 
\begin{align}
    \|\vec p - \vec q\|_1 \leq \epsilon = \sqrt{ \frac{2}{N} \ln \frac{2^m}{\delta}} \ , 
\end{align}
recovering the $\epsilon$ condition from \eqref{eq:bound_joint}. As in Appendix \ref{app:alfa-K}, with probability $1-\delta$ the actual state $\omega$ is in the feasible space and the bounds $\mathcal F_{\rm LB(UB)}$ hold. $\qed$

\section{Certification of state preparation $II$}
\begin{figure}[t]
	\centering
	\includegraphics[width=.5\linewidth]{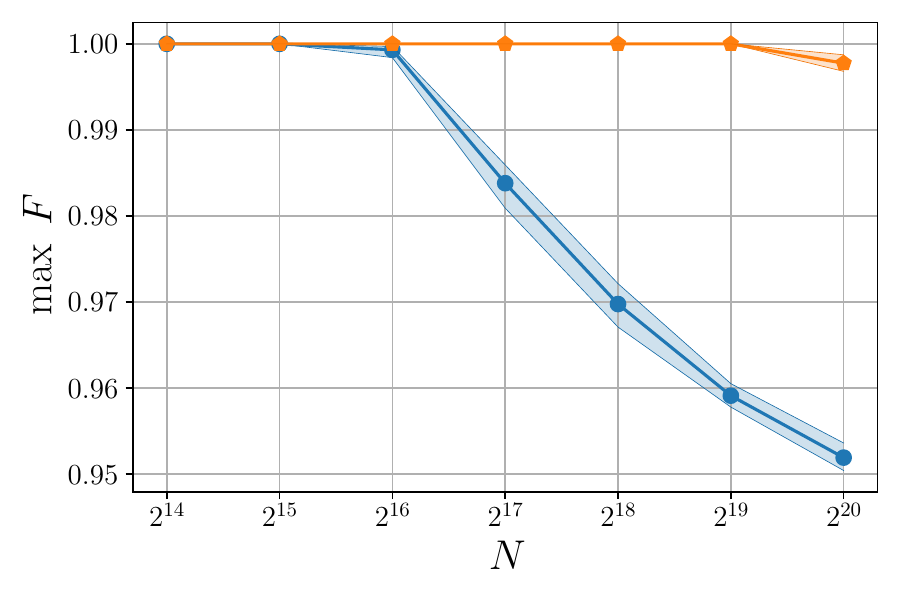}
	\caption{Maximum fidelity between a prepared state $0.9|\psi\rangle \langle \psi | + \frac{0.1}{16}I$ and the reference state $|\psi \rangle$ for $4$ qubits. The solid lines represent the median value obtained from $100$ Haar-random pure states using our method (blue dots) and method from Ref. \cite{de2023user} (orange pentagons), while the shaded areas depict the interquartile ranges. }
 \label{fig:max_fid}
\end{figure}

{For sake of illustration, we
present a last example concerning maximum fidelity of state preparation.}
In this application, we receive copies of a quantum state $\omega$, and we are told that this state corresponds to a pure state $| \psi\rangle$. We want to prove that this is false. To achieve this, we perform measurements on the given copies and determine the maximum fidelity $F(\rho) = \langle \psi | \rho | \psi \rangle$ between the states $\rho$ in the feasibility set $\Omega$ and the reference state $| \psi \rangle$. If the maximum fidelity differs from $1$, we can conclude that the state $\omega$ is not equal to $|\psi \rangle$. \\
In Fig \ref{fig:max_fid}, we prepare $100$ states of the form $0.9|\psi\rangle \langle \psi | + \frac{0.1}{16}I$, where $|\psi\rangle$ are Haar-random pure states of $4$ qubits. We subject these states to measurements using $16$ randomly chosen Pauli bases, with varying numbers of total shots. The joint-constraint protocol is then applied with confidence level of $0.997$. Blue dots depict the fidelity between the resulting states and the reference states $|\psi\rangle$. We compare these results with those obtained using protocol A from Ref. \cite{de2023user}, where the same confidence level and total shot count $N$ are used, but in this case, we measure the 81 Pauli strings that are informationally complete for 4 qubits. Our protocol needs around $2^{16}$ shots to succeed, while Protocol A needs around $2^{20}$, $16$ times more shots.\\

\end{document}